\documentclass[twocolumn,showpacs,preprintnumbers,amsmath,amssymb]{revtex4}
\usepackage{amsmath}
\usepackage{amsfonts}
\usepackage{amssymb}
\usepackage{epsfig}

\begin{document}

\preprint{}
\title{The optimization and shock waves in evolution dynamics}
\author{David B. Saakian$^{1,2}$}
\author{Jos\'e F. Fontanari$^{1}$}
 \affiliation{$^1$Instituto de F\'{\i}sica de S\~ao Carlos,
 Universidade de S\~ao Paulo, Caixa Postal 369, 13560-970 S\~ao Carlos, S\~ao Paulo, Brazil}
 \affiliation{$^2$Yerevan Physics Institute, Alikhanian Brothers St. 2, Yerevan 375036, Armenia }

\date{\today}

\begin{abstract}
We consider the optimal dynamics in the infinite population
evolution models with general symmetric fitness landscape. The
search of optimal evolution trajectories are complicated due to
sharp transitions (like shock waves) in evolution dynamics with
smooth fitness landscapes, which exist even in case of popular
quadratic fitness. We found exact analytical solutions for
discontinuous dynamics at the large genome length limit.
 We found the optimal
mutation rates for the fixed fitness landscape. The single peak
fitness landscape gives the fastest dynamics to send the vast
majority of the population from the initial sequence to the
neighborhood of the final sequence.
\end{abstract}
\pacs{ 87.23.Kg, 64.60.De} \maketitle

\medskip

\vskip 3 mm

\section{Introduction}
The search of optimization in evolutionary processes is a rather
popular idea in evolution research
\cite{fo99,fo00,tr07,ky07,su07,sc07,ch08}. Here we should
distinguish the optimization via mutation rate \cite{fo00} and via
fitness landscapes \cite{tr07}. The first is rather relevant for the
real biology. According to Darwinian view to evolution,  the force
driving evolution is the natural selection, not the creation of
genetic variants, because the mutations are random. Nevertheless
there have been experimental results which suggest that mutation
rates can vary, e.g. increasing during certain stresses \cite{fo99}.
This phenomenon has been called "adaptive mutation", which means
that the mutation rate is under the selective pressure \cite{fo00}.

 We will give the theory of optimal mutation
rates in case of infinite populations. The realistic case, of cause,
is connected with the finite population, and our results could be
considered just as a first step in that direction.

From the early days of Darwin-Wallace evolution theory [8] there has
been a hope that there is some optimization
 of the fitness during the evolution process.
Such picture really was confirmed in the case of proteins
\cite{ky07}.  In connection with [3] has been considered a
mathematical problem. What has been examined was a direct
parallelism with the famous Brachistochrone problem suggested in
1696 by Johann Bernoulli. Given two mutants, A and B, separated by
{\it n} mutational steps, what is the evolutionary trajectory which
allows a homogeneous infinite population of A to reach B in the
shortest time? In \cite{tr07} has been considered an approximate
solution of the mathematical problem in the case of finite
population, considering an optimization via fitness landscape. To
find an optimized evolution trajectory, one should have, first of
all, exact solutions of evolution dynamics. Such solutions have been
found only recently in the case of a single-peak fitness landscape
for the Crow-Kimura \cite{ck70,ba97,ba01} in \cite{sa04a} and for
the Eigen model \cite{ei71,ei89} in \cite{sa04b}. In [16] has been
solved the evolution dynamics (the manner of change of the mean
number of mutations in population)
 for the general symmetric
fitness landscape case, derived using the Hamilton-Jacobi equation
(HJE) method \cite{sa07,ka07}. The optimization problem is highly
complicated due to discontinuous effects in the dynamics of a mean
number of mutations, as has been found in \cite{sa08} (see also
\cite{ba01a,ca05}).
 The phenomenon exists even in smooth fitness
landscape. In Fig. 2 there is an example of such discontinuity. The
overlap $x^*$
 (mean number of mutations is defined as $1-2x^*/N$, N being the genome length)
 is first a smooth function of time, then its value
jumps from the low branch of the $S$ like loop to the upper one, and
again is a smooth function of time.
 As evolution equations
are exactly mapped into HJE equations, one can introduce
Hamiltonians and corresponding potentials. In \cite{sa08} has been
suggested a receipt to identify a class of discontinuities: when the
 evolution potential has two local maximums. It has been observed
also that the discontinuous dynamics could occur even for the case
of a single maximum in evolution potential, when the fitness
function is steep enough. The phenomenon is rather involved and  in
\cite{sa08} neither the "enough steepness" could be identified nor
the position
 of sharp transitions for any case of discontinuities.
 We will give an analytical
theory for some cases of discontinuous dynamics, and exact results
for the optimization via mutation rate. We found discontinuities
even in the quadratic fitness landscapes, missed in \cite{sa08}.

\section{The mutation optimization}

\subsection{Crow-Kimura model with asymmetric mutations}

 Consider the case
of model with different forward and back mutation rates as in
\cite{tr07}. For symmetric-fitness landscapes this model \cite{ba01}
assumes that the relative probabilities $p_l$, $l=0, 1, ..., N$ (N
being the genome length), obey
\begin{eqnarray}\label{e1}
\frac{d p_l}{dt} & = & p_l \left [ N f \left ( m_l \right ) -
\left ( \gamma_f - \gamma_b \right ) l + \gamma_b N  \right ] \nonumber\\
&  & + \gamma_b \left ( N-l+1 \right )p_{l-1} + \gamma_f \left ( l+1
\right ) p_{l+1},
\end{eqnarray}
where $m_l = 1 - 2 l/N$, and $p_l$ are relative probabilities at the
Hamming distance l (l mutations); $f(x)$ is a fitness function; and
$\gamma_f,\gamma_b$ are the mutation rates. In Eq.(1), for $l=0$ and
$l=N$ we omit $p_{-1}$ and $p_{N+1}$. Probability of having a
molecule at Hamming distance $l$ from the master is $p_l/\sum_kp_k$.
As in Ref.\cite{sa07}, at a discrete $x=1-2l/N$ we use the ansatz:
$p_l(t)\equiv p(x,t)\sim \exp[Nu(x,t)]$. Eq. (1) can be written as
Hamilton-Jacobi equation for $u\equiv \ln p(x,t)/N$ \cite{sa07}:
\begin{eqnarray}
\label{e2} \frac{\partial u}{\partial t}+H(x,u')\nonumber\\
-H(x,p)=f(x)-((1-x)\gamma_f+
\gamma_b(1+x))/2+\nonumber\\
\gamma_b\frac{1+x}{2}e^{2p}+\gamma_f\frac{1-x}{2}e^{-2p},
\end{eqnarray}
where $u' \equiv \partial u/\partial x$, the domain of $x$ is $
-1\le x\le 1$, and the initial distribution is $u(x,0)=u_0(x)$.
Minimizing $-H(x,p)$ via $p$, we get the expression of evolution
potential,
\begin{equation}\label{e3}
U(x)=f(x)+\sqrt{\gamma_b\gamma_f}
\sqrt{1-x^2}-\gamma_f\frac{1+x}{2}-\gamma_b\frac{1-x}{2}
\end{equation}
The evolution behavior is defined by the evolution potential
\cite{sa08}.

 We
solved Eq.(2) for $\gamma_f=\gamma_b=\gamma$ case in \cite{sa08} by
a method of characteristics \cite{me98,eva02}. For the
characteristics line $x(t)$ we have a Hamilton equation
$dx/dt=dH(x,p)/dp$. In our case Eq.(2) gives:
\begin{eqnarray}\label{4}
\dot x = \pm2\sqrt{k^2 -\gamma_f\gamma_b(1- x^2)},\nonumber\\
k\equiv q+\gamma_f\frac{1-x}{2}+\gamma_b\frac{1+x}{2}-f(x)
\end{eqnarray}
where $q\equiv \partial u(x,t)/\partial t$ is constant along the
characteristics, like the energy of the particle in classical
mechanics. At every point we have two characteristics, moving to the
right and left.

We consider the dynamics of the population in the Crow-Kimura model,
originally having fixed overlap $x_0$ with the reference (master)
sequence. Let us look at the manner of change in the mean overlap of
 the population $x^*(t^*)=\sum_jP_j(1-2d_i/N)$ at the
moment of time $t^*$.  $P_j$ is the fraction of the type $j$ in the
population, $d_j$ is the number of mutations in the $j$-th type
(compared with the master sequence), and such mutant has a fitness
$Nf(1-2d_i/N)$.
  As
time progresses the overlap distribution spreads out and so we focus
on the time evolution of overlap $x^*$ that yields the maximum of
this distribution.

Following to derivations of \cite{sa08}, we derived for the large
initial $x_0$
\begin{equation}\label{low_t_s}
t^* = \frac{1}{2}\int\limits_{x_0}^{x^*} d \xi ~ \left [ F \left (
\gamma, x^*,\xi \right ) \right ]^{-1/2}
\end{equation}
where we have the following expression for $F$:
\begin{eqnarray}\label{Fas}
F \left ( \gamma, x^*, \xi \right ) =  \left [ f \left ( x^* \right
)+ \gamma_f\frac{1+\xi}{2}+\gamma_b\frac{1-\xi}{2}- f \left (
\xi \right ) \right ]^2\nonumber\\
- \gamma_f\gamma_b \left ( 1- \xi^2 \right )
\end{eqnarray}
For the small $x_0$ there is another expression:
\begin{eqnarray}\label{high_t_s}
t^* &  = &  \frac{1}{2}  \int\limits_{x_0}^{x_1}
d \xi ~ \left [ F \left ( \gamma, x^*, \xi \right ) \right ]^{-1/2} \nonumber \\
& + &  \frac{1}{2}  \int\limits_{x^*}^{x_1} d \xi ~ \left [ F \left
( \gamma, x^*, \xi \right ) \right ]^{-1/2}
\end{eqnarray}
and $x_1$ is the solution of
\begin{equation}\label{def_x1}
F \left ( \gamma, x^*, x_1 \right ) = 0 .
\end{equation}
In \cite{sa08} has been considered the symmetric mutation scheme
$\gamma_f=\gamma_b=\gamma$ with $F_s$ instead of $F$:
\begin{equation}\label{Fs}
F_s \left ( \gamma, x^*, \xi \right ) =  \left [ f \left ( x^*
\right )+ \gamma- f \left ( \xi \right ) \right ]^2- \gamma^2 \left
( 1- \xi^2 \right )
\end{equation}

 For the quadratic fitness function
\begin{equation}
\label{f_x_2}
f \left ( x \right ) = \frac{c}{2} x^2
\end{equation}
Eqs. (5),(7) have real solutions provided that $x^* < 1 -
\gamma/c$. This upper bound determines the asymptotic value of the
overlap with the reference sequence. Of course, in the case
$\gamma/c > 1$ the selective phase is lost and the dynamics drifts
in the sequence space so that the asymptotic  regime is
characterized by a zero overlap with the reference sequence. To
decide which equation to use we need to calculate $t_h$
\begin{equation}\label{Th_s}
t_h = \frac{1}{2} \int \limits_{x_h}^{x_0} d \xi ~ \left [ F_s \left
( \gamma, x_h, \xi \right ) \right ]^{-1/2}
\end{equation}
where $x_h$ is a root of $F_s \left ( \gamma, x_h, x_0  \right )=
0$. This equation has a solution provided that $f \left ( x_h \right
) \leq f \left ( x_0 \right )$, which in the case of monotonically
increasing fitness implies $x_h \leq x_0$. Thus for a given $x_0$
and $t^*$ we calculate $x_h$ and then $t_h$. If $t^* < t_h$ we use
Eq.\ (\ref{low_t_s}), otherwise we use Eq. (8), to obtain $x^* = x^*
\left ( t^* \right )$.

\subsection{ Discontinuous dynamics in case of quadratic fitness
function}
\begin{figure}
\centerline{\epsfig{width=0.6\columnwidth ,file=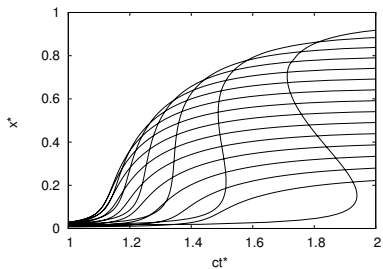}}
\par
\caption{The dynamics of $x^*(t^*)$ (most likely value of the
overlap with the reference sequence as function of time $t^*$) by
Eq. (7) for the symmetric case $\gamma_f = \gamma_b = \gamma$ and
(top to bottom at $ct^* =2$) $\gamma/c = 0.05,0.1, \ldots,0.7,0.75$.
The initial population has overlap $x_0 = 0.01$ with the reference
sequence. For $t^* \to \infty$ we find $x^* = 1 -  \gamma/c$. }
\label{DYN:1}
\end{figure}

\begin{figure}
\centerline{\epsfig{width=0.6\columnwidth ,file=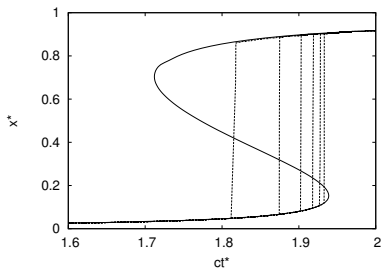}}
\par
\caption{Numerical solution of the ODE system (7) for $x_0 = 0.01$,
$\gamma/c = 0.05$ and (dashed vertical lines from left to right)
$N=2000, 4000, \ldots, 12000$.  For $N \to \infty $ the  jump in
$x^*$ takes place at $ct^* = ct_d^* = 1.939$ and has size $\Delta
x^* = 0.755$. } \label{DYN:2}
\end{figure}

\begin{figure}
\centerline{\epsfig{width=0.6\columnwidth
,file=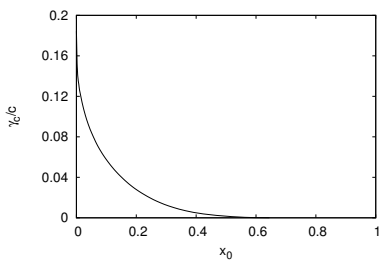}}
\par
\caption{The critical line $\gamma_c/c$ vs. $x_0$ at which $\Delta
x^* = 0$. The critical $c$ is defined from the system of equations
$dx^*/dt^*=0,d^2x^*/d^2t^*=0$. Below this curve the most probable
overlap $x^*$ undergoes a discontinuous transition at $t^* = t_d^*$
(see Fig.\ \ref{DYN:2}). } \label{DYN:4}
\end{figure}
\begin{figure}
\centerline{\epsfig{width=0.6\columnwidth ,file=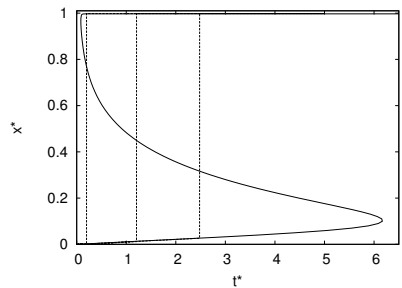}}
\par
\caption{ The relaxation from the original flat distribution with
the fitness function $f(m)=4*
exp(8(m-1)),\gamma=0.1,N=1000,5000,10000$. The jump is at the point
$dx^*/dt^*$ at $N=\infty$.} \label{DYN:5}
\end{figure}

In \cite{sa08} there have been derived analytical formulas
Eqs.(5),(7) with $F=F_s$. The \cite{sa08} failed to describe the
discontinuities of $x^*(t)$ analytically.
 The mean fitness
is defined as a minimum of
 $U(x)$.
When this function has two maxima at $1\ge x>0$, there is a
discontinuity in the dynamics, \cite{sa08}. The point is that there
can be singularities in the dynamics, even for the fitness with a
single maximum at $x>0$, when the fitness is too steep. We performed
a numerics  for symmetric mutations ($\gamma_f=\gamma_b=\gamma$) to
clarify the character of discontinuous dynamics, see Fig.1-Fig.3. In
the selective phase $c> \gamma$, the potential $U(x)$ has a single
maximum at $1>x>0$. Nevertheless, sometimes the function $x^*(t^*)$
has jumps.

 Figure \ref{DYN:1} illustrates the time
evolution of $x^*$ for $x_0 = 0.01$. For not too small $\gamma/c$
the  $x^*(t^*)$ is a monotonic function, and the direct numerics of
the system of equation for Crow-Kimura model supports well the
theoretical formulas for $x^*(t^*)$.

 For small values of $\gamma/c$
the S-shaped curves indicate the existence of a discontinuity in the
position of the maximum of the overlap probability distribution.
This threshold phenomenon was overlooked in a previous analysis of
this problem which considered a single parameter setting, $c=2$ and
$\gamma = 1$ \cite{sa08}.

The unusual time dependence of $x^*$ exhibited in Fig.\ \ref{DYN:1}
is  quite counter-intuitive since it implies that for, say,
$\gamma/c = 0.05$ there is an entire range of overlap values which
are never reached by the evolutionary dynamics. To check that
finding and to gather information on the stability of the solutions
in the multi-solution regime,  we present in Fig. 2 the results of
the numerical solution of the ODE system  (1) for different values
of sequence lengths. These results not only confirm the theoretical
predictions but complement them by showing that the solution
corresponding to the lower branch of the S-shape is the stable one.
This information allows us to obtain the value $t^* = t_d^*$ at
which the discontinuity takes place as well as the size of the
discontinuity $\Delta x^*$. This can be done by locating the lower
value of $x^*$ for which $dt^*/dx^* = 0$ in Eq.\ (7).

Our conclusion, deduced from the analysis of Fig. 2 that the jump in
the dynamics occurs at the point where $dx^*/dt^*=0$, is a rather
general one. We checked that it is valid in other cases with
discontinuous dynamics as well, see Fig. 4.


\subsection{ Optimal mutation rates in case of fixed overlap value in
original population}
 Here we explore  another important result
exhibited in Fig.\ \ref{DYN:1}, namely, that there is an optimal
value of the scaled mutation rate $\gamma/c$ that minimizes the
evolutionary time to go from $x_0$ to $x^* > x_0$.  To assess this
point in more detail, we note first  the obvious fact that this
evolutionary trajectory is possible only for $\gamma/c < 1 - x^*$.
With this fact in mind, we can see from Fig.\ \ref{DYN:1} that to
reach the end point, say, $x^* = 0.2$ it is a bad strategy to choose
either small or large values of $\gamma/c$. In fact, there is an
optimal value of the mutation rate, which for the parameter setting
of this example ($x_0 = 0.01$ and $x^* = 0.2$) is $\gamma_{opt}/c =
0.3632$.  This interesting analytical finding  substantiates the
empirical strategy of fine tuning the mutation rate in Genetic
Algorithms \cite{GA}.

\begin{figure}
\centerline{\epsfig{width=0.6\columnwidth
 ,file=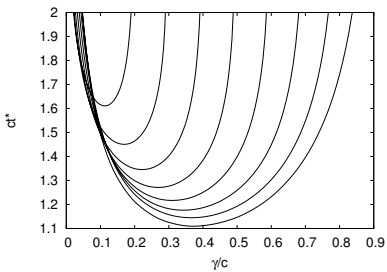}}
\par
\caption{Time period $t^*$ needed for the maximum of the overlap
distribution  to reach the values (right to left) $x^* = 0.1,0.2,
\ldots, 0.8$ as function of  $\gamma/c$.  The initial population has
overlap $x_0 = 0.01$ with the reference sequence. The dynamics can
reach $x^*$ provided that $\gamma/c < 1 - x^*$. } \label{OPT:1}
\end{figure}

Figure  \ref{OPT:1} neatly illustrates the existence of an optimal
mutation rate for the fixed initial condition $x_0 = 0.01$.  To draw
one of the curves in this figure we keep the end point $x^*$ fixed
and  measure the evolution time as a function of the scaled mutation
rate. The existence of an optimal mutation rate that corresponds to
the fastest evolutionary trajectory (i.e., minimum $t^*$) for the
particular fitness choice, Eq.\ (\ref{f_x_2}), is patent from this
figure.

To find the exact location of the minima exhibited in Fig.\
\ref{OPT:1} we put the condition $\frac{dt^*}{d \gamma} =0$, and get
\begin{eqnarray}\label{dtdg}
 - \frac{1}{4} \int \limits_{x_0}^{x_1}
\frac{d \xi}{F_s^{3/2}}  \frac{dF_s(\gamma,x^*,\xi)}{d\gamma} -
\frac{1}{4} \int \limits_{x^*}^{x_1} \frac{d \xi}{F_s^{3/2}}
\frac{dF_s(\gamma,x^*,\xi)}{d\gamma} =\nonumber\\
\frac{1}{\epsilon^{1/2}}
\frac{\frac{dF_s(\gamma,x^*,\xi)}{d\gamma}}{\frac{dF_s(\gamma,x^*,\xi)}{dx_1}}\nonumber\\
F_s \left ( \gamma, x^*, x_1 \right ) = \epsilon
\end{eqnarray}
 Here $\epsilon$ must be set to a small (but not too
small) value, typically $\epsilon = 10^{-6}$.
 We also simply
opted for the direct numerical derivation of the curves  shown in
Fig.\ \ref{OPT:1}.
 What is surprising is that the
optimal mutation rate $\gamma_{opt}$ grows very steeply as $x^*$
departs from $x_0$ and quickly reaches a maximum value. Looking
carefully the Eq. (7), we see an important issue. The optimization
depends on the behavior of the fitness function outside the interval
$[x_0,x^*]$.


\subsection{ Originally flat distribution}

\begin{figure}
\centerline{\epsfig{width=0.6\columnwidth ,file=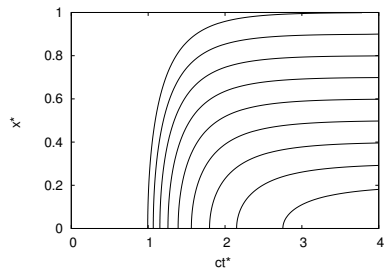}}
\par
\caption{ Location $x^*$ of the maximum of the overlap distribution
as function of time $t^*$ for the symmetric case $\gamma_f =
\gamma_b = \gamma$ and (top to bottom) $\gamma/c = 0, 0.1, 0.2,
\ldots, 0.8$. The initial population is uniformly distributed in
sequence space. For $t^* \to \infty$ we find $x^* = 1 -  \gamma/c$.
} \label{DYN:5}
\end{figure}

 For the initially flat
distribution we have \cite{sa08}
\begin{equation}\label{e13}
u(0, x) =  -\frac{1+x}{2} \ln \frac{1+x}{2}- \frac{1-x}{2} \ln
\frac{1-x}{2} ,
\end{equation}
and
\begin{equation}\label{t_uni_s}
t^* = \frac{1}{2}\int\limits_{x^*}^{x_1} d \xi ~ \left [ F_s \left (
\gamma, x^*,\xi \right ) \right ]^{-1/2}
\end{equation}
where $F_s$ and $x_1$ are defined by Eqs.\ (\ref{Fs}) and
(\ref{def_x1}), respectively. As pointed out in Ref.\ \cite{sa08},
the initial overlap distribution has a peak at $x=0$, which yields
the maximum of the overlap distribution for $t < t_0$ where
\cite{sa08}
\begin{equation}\label{t_0_flat}
 ct_0 = \frac{ \cos^{-1} \left ( 1 - \gamma/c \right
)^{1/2}} {\left [ \gamma/c \left ( 1 - \gamma/c \right ) \right
]^{1/2}} .
\end{equation}
Note that $ct_0 \to 1$ for $\gamma/c \to 0$, and $ct_0 \approx
\sqrt{2} \left ( 1 - \gamma/c \right )^{-1/2}$ for $\gamma/c \to 1$.
We turn now to the case where the initial population is uniformly
distributed among the $2^N$ configurations.

 Figure \ref{DYN:5} shows the time evolution of $x^*$ for the
flat initial  distribution. The results are in stark contrast with
those of the peaked initial distribution (see Fig.\ \ref{DYN:1}):
the odd S-shaped curves that produced the interesting dynamic
behavior discussed before are absent in this case. In addition, the
curves for different values of  $\gamma/c$  never cross which
indicates that the fastest trajectory to reach any point $x^*$ is
given by a vanishingly small mutation rate.

Let us calculate the optimal period to have a peak of population
with the overlap $x^*$ at the moment of time $t^*$. We should find
the $x^*(t^*)$ looking the maximum of the
\begin{eqnarray}\label{e16}
 -\frac{1+x^*}{2} \ln \frac{1+x^*}{2}-
\frac{1-x^*}{2} \ln \frac{1-x^*}{2}+t^*f(x^*)
\end{eqnarray}
\section{ Fitness Optimization}
 Although the selection of a fitness function that
minimizes the evolution time between any two points $x_0$ and $x^*$,
which correspond to the maximum of the  overlap distribution in two
distinct times,  is not as biologically significant as the selection
of the optimal mutation rate, it has a considerable aesthetical
appeal as the problem is somewhat akin to the Brachistochrone
problem of physics \cite{tr07}. In \cite{tr07} has been assumed that
the fastest evolution dynamics between two sequences  is given by a
single peak fitness. The point is that one should accurately
formulate the optimization task. The first possibility- we look the
arrival of some fraction of population to the master peak. The
second version: we look the arrival of a vast majority of population
to the small (the Hamming distance is miserable compared with N)
neighborhood of the master sequence. The situation is highly
non-trivial. If we took the first version with some small fraction,
then the linear fitness could give better results than the
single-peak fitness, see the Fig. 7.

 If we take the second version
of optimization, then the single-peak fitness looks like as the
fastest one.
 For the considered case (from sequence to sequence) we can
just give the expression of the minimal time following to the
results by \cite{sa04b}.

For the symmetric mutation case the fastest relaxation gives the
single-peak fitness landscape ($r_0=J$ for the peak sequence and
$r_i=0$ for the other sequences). In
 \cite{sa04b} has been found the relaxation period to send the
 population from the given sequence (at the Hamming distance $N(1-m)/2$ from the peak sequence) to the peak
 one.
To find the minimal time $t$ we just add the optimization condition
via the choice of $\gamma$ to the solution of \cite{sa04a}:
\begin{eqnarray}
 \label{e17}
t=\frac{\phi(x,t_1)-Jt_1}{J-\gamma}\nonumber\\
 \frac{1+x}{2}\tanh(\gamma t_1)+\frac{1-x}{2\tanh(\gamma t_1)}
 -\frac{J}{\gamma}=0\nonumber\\
 \phi(x,t)=[\frac{1+x}{2}\ln\cosh (\gamma
t)+\frac{1-x}{2}\ln\sinh(\gamma t)]\nonumber\\
\frac{\partial \phi}{\partial \gamma}=0
\end{eqnarray}
We have done some numerics, see Fig. 7, supporting the choice of
single-peak fitness as an optimal fitness for the fastest
relaxation, and $t$ by Eq. (17) as a minimal time period.


Consider now the fitness optimization problem in case of overlap
distributions (to send the population from the original overlap with
$m=x_0$ to the eventual one with $m=x^*$) and symmetric fitness
landscape.
 We are looking the
optimization problem for  the special fitness with
\begin{eqnarray}\label{18}
f(m)=0,x<x^*\nonumber\\
f(x^*)=J
\end{eqnarray}
Eq.(8) gives $x_1=x^*$, then  Eq. (7) is simplified: the first term
disappears. For the fitness by Eq.(18) we have
\begin{eqnarray}\label{e19}
t^*   =   \frac{1}{2}  \int\limits_{x_0}^{x^*} \frac{d \xi}{ \sqrt{(
J+ \gamma )^2- \gamma^2 ( 1- \xi^2)}}
\end{eqnarray}
It is easy to check that the minimal time is given by the  fitness
of Eq.(18). As
\begin{eqnarray}\label{e20}
 \sqrt{(J+ \gamma )^2- \gamma^2 ( 1- \xi^2)}>
 \sqrt{(J+ \gamma- f (\xi ))^2- \gamma^2  ( 1- \xi^2 )}
\end{eqnarray}
the time given by Eq.(19) is less than the time given by Eq.(7) for
any $f(m)>0$.

\begin{figure}
\centerline{\epsfig{width=0.6\columnwidth,file=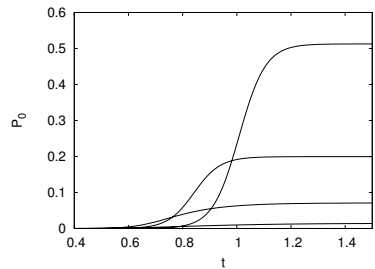}}
\par
\caption{ Location These are the results of the simulations for
$N=20$, symmetric mutation rate $\gamma = 1$. Fitness $f(x) = 2 N
x^a$ for $a=1,2,4$ (show in the figure) SP $f(x) = 0$ for $x < 1$
and $f(1) = 2 N$.
 } \label{sym:1}
\end{figure}

\section{ Directed mutation case} Consider the case of asymmetric
mutations \cite{tr07}. We have original distribution at some $x_0$,
and our goal is to send the population to the overlap $x^*$. Now
there is a single characteristic. therefore, contrary to the
symmetric mutation case, all the properties are defined via the
behavior of the fitness function in the considered interval
$[x_0,x^*]$.
 Eqs.(4),(5) give the
following equation
\begin{eqnarray}
\label{e21} t^*=\frac{1}{2}\int_{x_0}^{x^*} \frac{dx}{f(x_*)+\gamma
\frac{1-x}{2}-f(x)}
\end{eqnarray}
We see that the optimization via mutation is trivial: raising the
mutation rate we can send the population  to the point $x^*$
immediately. The optimization via fitness is also trivial: the
fastest trajectory is via the fitness $f(m)=0,m<m^*$ and $f(1)=J_0$.

We have done a numerics for quadratic fitness case, see Fig. 8. We
see that the results again support the conjecture that the jumps are
at the point with $dx^*/dt^*=0$.
\begin{figure}
\centerline{\epsfig{width=0.6\columnwidth ,file=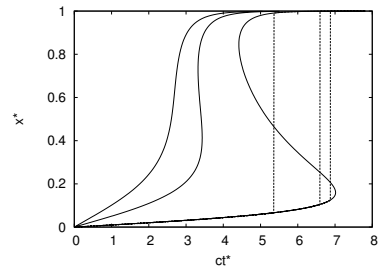}}
\par
\caption{ Location $x^*$ of the maximum of the overlap distribution
as function of time $t^*$ for directed mutation case with $x_0 = 0$,
(left to right) $\gamma_f/c = 0.1, 0.05$ and $0.01$. The numerical
solution of the system (1) is given by the  dashed vertical lines
for (left to right) $N=1000,5000$ and $10000$. } \label{ASY:1}
\end{figure}
\section{ Discussion}
We considered the problem of optimization in evolution in case of
infinite population, symmetric fitness landscape and large genome
length and found exact solutions. We found that the optimization
(optimal control, see \cite{fl75}) is a highly non-trivial problem,
as sharp, discontinuous transitions are typical for the evolution
dynamics even with smooth fitness landscapes. We investigated these
discontinuities and  gave an analytical description of such sharp
transitions for symmetric smooth landscape. We could succeed doing
numerics for a larger values of N than those in  \cite{sa08}. We
found dynamical discontinuities even in case of directed mutations.
The sharp transitions in evolution are important regarding the
punctual evolution phenomenon (see \cite{go93} and the review
\cite{dr01}).

The optimization via mutation rate is most intriguing from the point
of view of adaptive mutations. We calculated the minimal time to
send the population from original sequence with small overlap (with
the master sequence) and low fitness  to the some final one (with a
higher fitness). The solution of the optimization problem is
nontrivial, and there is some optimal mutation rate. It is
interesting that the optimal rate of mutation to send the population
from the overlap $x_0$ to $x^*$ is defined with the behavior of the
fitness outside the interval $[x_0,x^*]$.
 The numerics confirm
our analytical results. On the contrary, when we need to send the
population from the high fitness configurations with the fixed
original overlap to the final one with lower overlap, the mutation's
optimization is a trivial task: just increase the mutation rate.
Similar is the situation in case of directed mutation: one can send
the population in a fastest way just increasing the mutation rate.

If we consider the evolution from originally flat distribution (all
sequences have the same probability), then the optimal mutation rate
is zero.

The optimization via fitness landscape (to send the population from
the original sequence to the final sequence) should be carefully
defined as a mathematical problem. When we are interested to send
some small fraction of the population to the master sequence, the
linear fitness can give better results than the single-peak one.
When we are looking how to send the vast majority of population to
the some infinitesimal neighborhood of the master sequence, then the
optimum
 is given by a single-peak fitness
landscape. Such hypothesis has been assumed first in \cite{tr07}. We
could not prove it rigorously, but we gave just an exact expression
for  this optimal time period, as well as performed numerics
illustrating the optimization. If we are looking how to send the
population with the initial overlap $x_0$ to the final overlap
$x^*$, then, as we proved rigorously, the minimal time is given by
the single-peak like fitness Eq. (18).

We looked only at the infinite population problem. Our consideration
could be a first step in consideration to the real biological
situation.

The work at Yerevan was supported in part by the VolkswagenStiftung
grant ``Quantum Thermodynamics''. The research at S\~ao Carlos was
supported in part by CNPq and FAPESP, Project No. 04/06156-3. D.B.S.
thanks to the hospitality of the Instituto de F\'{\i}sica de S\~ao
Carlos, Universidade de S\~ao Paulo, and the FAPESP travel grant No.
08/10420-9 for  the support to his visit to S\~ao Carlos.


\end{document}